\documentclass[10pt,twocolumn,letterpaper]{article}

\usepackage[pagenumbers]{cvpr}      %

\ifdefined\pdfsuppressptexinfo\pdfsuppressptexinfo=-1\fi  %
\usepackage{cuted}      %
\usepackage{currfile}
\usepackage{colortbl}   %
\usepackage{multirow}
\usepackage{bm}

\xspaceaddexceptions{\% \, ] $}
\newcommand{\sNcases}{1,328\xspace}
\newcommand{\sNdsa}{1,982\xspace}
\newcommand{\sNfd}{1,054\xspace}
\newcommand{\sNproc}{610\xspace}
\newcommand{\sNrunsEval}{1,701\xspace}
\newcommand{\sNrunsExt}{346\xspace}
\newcommand{\sNrunsInt}{270\xspace}
\newcommand{\sNtestExt}{122\xspace}
\newcommand{\sNtestInt}{98\xspace}
\newcommand{\sNtrain}{342\xspace}
\newcommand{\sNval}{48\xspace}
\newcommand{\sNweb}{274\xspace}

\newcommand{\method}{X-Strata\xspace}

\definecolor{oursbg}{HTML}{EAF2FA}
\definecolor{cA}{HTML}{6B6B6B}
\definecolor{cT}{HTML}{E69F00}
\definecolor{cP}{HTML}{2A9D8F}
\definecolor{cX}{HTML}{0072B2}
\definecolor{cS}{HTML}{CC79A7}
\newcommand{\cueon}[1]{\textcolor{#1}{\rule{0.56em}{0.56em}}\hspace{0.12em}}
\newcommand{\cueoff}{\textcolor{black!17}{\rule{0.56em}{0.56em}}\hspace{0.12em}}
\newcommand{\best}[1]{\textbf{#1}}
\newcommand{\second}[1]{\underline{#1}}

\newcommand{\vu}{\mathbf{u}}

\newcommand{\teasergraphic}{%
  \includegraphics[width=\linewidth]{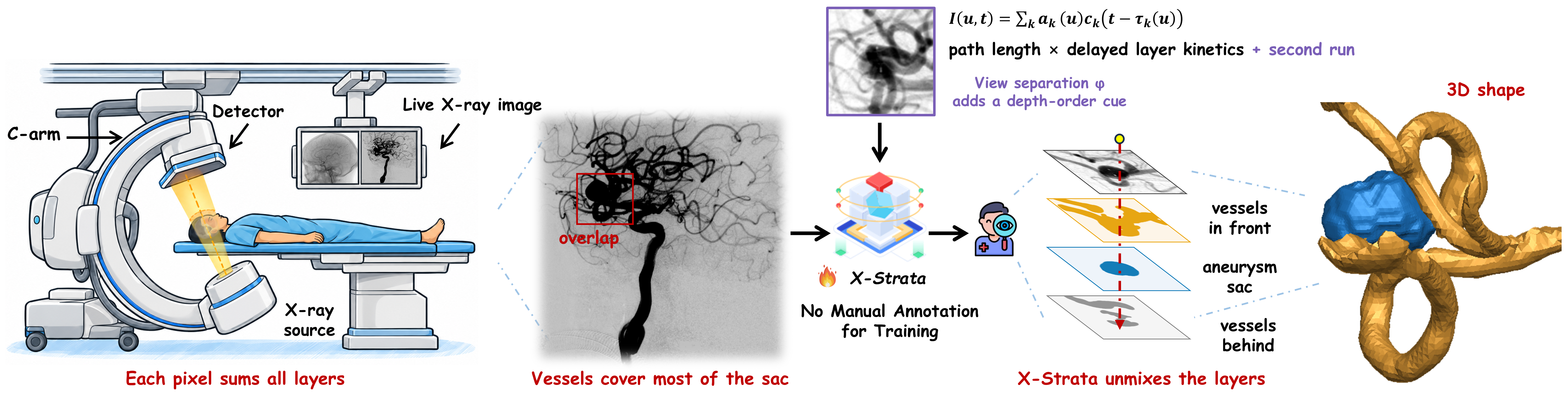}%
  \llap{%
    \setlength{\unitlength}{\linewidth}%
    \begin{picture}(1,0)
      \put(0.50789,0.03297){\color{white}\rule{0.11966\linewidth}{0.02765\linewidth}}
    \end{picture}%
  }%
}

\renewcommand{\paragraph}[1]{\vspace{0.35em}\noindent\textbf{#1}\hspace{0.45em}}

\makeatletter
\@ifundefined{leftlinenumbers}{}{%
  \leftlinenumbers
  \setlength{\linenumbersep}{5pt}%
}
\makeatother

\usepackage{microtype}
\usepackage{placeins}

\definecolor{hlres}{HTML}{FFF0A0}
\definecolor{hlauth}{HTML}{FFD3A6}
\newif\ifshowhl \showhlfalse

\newcommand{\newIntODDelta}{1.45\xspace}
\newcommand{\newIntODLo}{-0.50\xspace}
\newcommand{\newIntODHi}{3.37\xspace}

\newcommand{\newIntDOADelta}{1.25\xspace}
\newcommand{\newIntDOALo}{0.01\xspace}
\newcommand{\newIntDOAHi}{2.53\xspace}
\newcommand{\newIntVol}{17.21\xspace}

\newcommand{\newExtODDelta}{1.11\xspace}
\newcommand{\newExtODLo}{-0.63\xspace}
\newcommand{\newExtODHi}{2.83\xspace}

\newcommand{\newExtDOADelta}{0.63\xspace}
\newcommand{\newExtDOALo}{-0.57\xspace}
\newcommand{\newExtDOAHi}{1.89\xspace}
\newcommand{\newExtVol}{25.81\xspace}
\newcommand{\newFullECE}{0.030\xspace}
\newcommand{\newSingleECE}{0.073\xspace}
\newcommand{\newTTP}{0.144\xspace}

\definecolor{cvprblue}{rgb}{0.21,0.49,0.74}
\usepackage[pagebackref,breaklinks,colorlinks,allcolors=cvprblue]{hyperref}

\def\confName{}
\def\confYear{}

\title{X-Strata: Separating Aneurysms and Overlapping Vessels into Layers from 2D X-ray Angiography}

\author{Zhihong Wen$^{1}$\quad Congzhou Chen$^{1}$\quad Hucheng Li$^{2}$\quad Xun Han$^{3}$\quad Jianning Zhang$^{3}$\quad Yakun Chen$^{3}$\\
Hao Gao$^{3}$\quad Hao Zhao$^{4}$\quad Yudong Ma$^{5}$\quad Guanqiao Li$^{1,*}$\quad Yongping Liang$^{3,*}$\quad Shengli Guo$^{3,*}$\\[3pt]
{\small $^{1}$Vanke School of Public Health, Tsinghua University, Beijing, China}\\
{\small $^{2}$College of Computer Science and Electronic Engineering, Hunan University, Changsha, China}\\
{\small $^{3}$Department of Neurosurgery, the First Medical Center of Chinese PLA General Hospital, Beijing, China}\\
{\small $^{4}$Department of Neurosurgery, the Sixth Medical Center of Chinese PLA General Hospital, Beijing, China}\\
{\small $^{5}$Department of Neurosurgery, Air Force Medical Center of the Chinese PLA, Beijing, China}\\
{\small $^{*}$Co-corresponding authors}}

\begin{document}
\maketitle
\setlength{\stripsep}{4pt plus 1pt minus 1pt}
\begin{strip}
\centering
\teasergraphic
\captionsetup{hypcap=false}
\captionof{figure}{\textbf{Layer separation in X-ray angiography.}
Each log-subtracted pixel integrates contrast attenuation along a ray.
The left panel shows the C-arm acquisition setup.
The center panel marks sac--vessel overlap on a peak digital subtraction angiography frame.
The right panel illustrates the target layers, which comprise the amodal sac and the vessels in front of and behind it.
Parallax between the two views provides a cue to their depth order.}
\captionsetup{hypcap=true}
\label{fig:teaser}
\end{strip}

\begin{abstract}
An X-ray angiogram sums every structure along a ray into one pixel. When vessels cross an aneurysm, they obscure part of it and mix their contrast into its filling curve. We present X-Strata, a network that separates a cine angiogram into three layers: the complete aneurysm sac, the vessels in front of it, and the vessels behind it. Contrast timing indicates how much each layer contributes to a pixel, and parallax between two views indicates their order, which a single view cannot determine. X-Strata first estimates the layers and their contrast curves, then uses the second view to assign vessels to the front or back layer, and lifts all layers into 3D. It learns from layer labels projected from same-session 3D rotational angiography, but needs only 2D cine runs and their acquisition geometry at inference. On 220 held-out patients from three centers (122 external), two-view X-Strata ranks first among 14 baselines, including VGGT and C$^2$RV, in recovering the sac where vessels overlap it (Dice 0.693 internal, 0.641 external) and in ordering the layers (balanced accuracy 0.703, 0.659). Its order accuracy exceeds that of all 11 end-to-end baselines by at least 4.1 points internally and 3.3 externally (Holm-adjusted \(p < 0.001\)) and grows with view separation. Recovered 3D sac surfaces have a median Chamfer distance of 0.34 mm on the internal cohort. X-Strata may therefore provide layered aneurysm and vessel information, including 3D sac shape, for patients who undergo only 2D angiography.
\par
\end{abstract}

\section{Introduction}
\label{sec:intro}

In digital subtraction angiography (DSA), each pixel integrates contrast attenuation along its ray~\cite{auvray2005,fischer2015}, so an intracranial aneurysm and the vessels crossing it collapse into one image (\cref{fig:teaser}). The overlap obscures part of the aneurysm sac and mixes vessel contrast into the curves used to characterize washout~\cite{sadasivan2002,golitz2020} and to build parametric maps~\cite{shields2024,bhurwani2020,mondal2024}. 3D rotational angiography (3D-RA) can resolve the overlap, but not every patient undergoes it. We therefore ask whether routine 2D cine runs can be separated into their layers.

Unlike reflections and occluders in natural images~\cite{reflection2018,seethrough2020,omnimatte2021,lna2021,pix2gestalt2024,dvas2025}, the layers of an angiogram are both additive and ordered along the ray. Contrast reaches the sac and the vessels at different times, so their time courses indicate how much each contributes to a pixel. The order is harder: exchanging two layers along a ray preserves the cine if their path lengths, contrast curves, and arrival delays stay fixed, so a single parallel view cannot determine it. A second view adds parallax.

Existing methods address parts of this problem. Segmentation and amodal completion recover masks but neither a layer's contribution nor its order. Vessel stereo estimates depth~\cite{coste1999,liao2010} but inherits its masks and leaves the contrast unmixed. Sparse-view reconstruction~\cite{difnet2023,c2rv2024} and feed-forward 3D models~\cite{vggt2025} can supply every layer after projection, but they do not separate layers by contrast kinetics.

We present \method, a network that separates cine DSA into three layers: the amodal aneurysm sac, the vessels in front of it, and the vessels behind it (\cref{fig:method}). Our key idea is to use each cue for what it measures: contrast kinetics for how much each layer contributes, and parallax for the order of the layers. A first pass estimates layer maps and contrast curves, a second pass searches for vessel depth along centerlines in the second view, and the layers are lifted into a shared 3D representation. Same-session 3D-RA supplies projected labels during training, and inference uses only 2D cine runs and their geometry.

We make three contributions: an analysis of what cine DSA can determine about overlapping layers (\cref{sec:analysis}); a model that estimates per-layer path lengths, contrast curves, and order probabilities from 2D cine runs; and an evaluation on 220 held-out patients from three centers, 122 of them external. With two views, \method ranks first among 14 baselines on overlap-region sac Dice and balanced depth-order accuracy in both cohorts, and its order accuracy exceeds that of all 11 end-to-end baselines by at least 4.1 and 3.3 points (Holm-adjusted $p<0.001$). As our analysis predicts, order accuracy stays near 0.5 for every one-run method and rises with view separation, to 0.75 and 0.71 at wide angles. Recovered 3D sac surfaces have median Chamfer distances of 0.34 and 0.48\,mm. These results suggest that layered aneurysm and vessel anatomy, including 3D sac shape, may be recoverable for patients who undergo only 2D angiography.

\section{Related work}
\label{sec:related}

\paragraph{Image layers and amodal support.}
Reflection separation exploits gradient, motion, and coupled image priors~\cite{reflection2018,seethrough2020,doubledip2019} or the motion and decorrelation of additive layers~\cite{szeliski2000,sarel2004}. Omnimatte assigns objects to RGBA layers~\cite{omnimatte2021,lna2021,omnimatterf2023}, and amodal segmentation infers full object support from shape priors, occluder-aware transformers, or generative completion~\cite{xiao2021,c2fseg2023,aisformer2022,pix2gestalt2024,dvas2025}. Angiographic layers are additive like reflections and ordered along the ray like occluders, so contrast changes over time can constrain the covered part of the sac.

\paragraph{Temporal decomposition.}
The sum of layer path lengths weighted by contrast curves has the structure of a nonnegative factorization~\cite{leeseung1999}. Under separability, pure samples help identify the source curves~\cite{donoho2003,arora2012,gillis2014,fu2019}, and related factor models are used in dynamic imaging, calcium imaging, and hyperspectral unmixing~\cite{barber1980,sitek2002,chen2011camcm,cnmf2016,bioucas2012}. Even known curves can be difficult to separate when they are nearly collinear. Factorization alone also cannot determine depth order.

\paragraph{Angiographic layers and labels.}
Layer decomposition in radiographs and fluoroscopy has used CT-supervised depth slabs, bone separation, motion, low-rank structure, and canonical layers~\cite{albarqouni2017,blsgan2025,auvray2005,preston2013,fischer2015,ma2017layer,jin2017rpca,denver2025}. Single-view vessel-depth channels have also been proposed~\cite{siemenspatent}, although a single parallel view leaves depth order ambiguous (\cref{sec:analysis}). DSA segmentation and parametric imaging use visible intensity and temporal signals~\cite{jin2020dsa,liu2025dsa,dias2024,hong2019ica,bhurwani2020,shields2024,mondal2024}. Projected CT labels have supported fluoroscopic segmentation of static anatomy~\cite{grupp2020,fluorosam2024}, whereas here 3D-RA labels the projected vascular layers and the cine supplies the contrast dynamics of its own injection.

\paragraph{Two-view geometry.}
Multi-view vessel reconstruction has used epipolar matching with path search and graph cuts~\cite{coste1999,liao2010,cimen2016}. Shape estimation from two projections has used contour morphing and ellipse fitting~\cite{raghavan2014,williams2025mp}. A one-to-one match or fixed ordering can constrain opaque stereo~\cite{ohta1985}, whereas a ray through a transparent scene can intersect several structures~\cite{tsin2006,heitz2010}. Epipolar consistency relates the projected mass of two transmission views~\cite{aichert2015,wurfl2019,rohleder2023}, and we use the first moments of each layer as a depth reference. Stereo solvers take fixed vessel masks as input, whereas \method revises its layer maps while inferring depth.

\paragraph{Reconstruction from few views.}
Per-scene and feed-forward methods recover attenuation or another 3D representation from sparse projections~\cite{naf2022,saxnerf2024,xgaussian2024,r2gaussian2024,x2ctgan2019,difnet2023,c2rv2024,xlrm2026,xgrm2025,deepca2025} but do not model contrast dynamics. Time-resolved methods model contrast flow or cardiac motion but optimize a 4D volume per scene~\cite{tiavox2024,fourdrgs2025,nerfca2024}. Closest to our setting, Cafaro \etal reconstruct cerebral vessels from two projections~\cite{cafaro2024}, Frisken \etal add centerline and contrast-flow constraints to biplane reconstruction~\cite{frisken2025}, and Williams \etal recover 4D angiography from two views given a prior 3D geometry~\cite{williams2025}. These methods output vessel volumes. \method instead outputs the amodal sac and the vessels in front of and behind it, together with their contrast curves, and needs 3D data only for training.

\section{What an angiogram reveals about layers}
\label{sec:analysis}
\begin{figure*}[t]
\centering
\includegraphics[width=\linewidth]{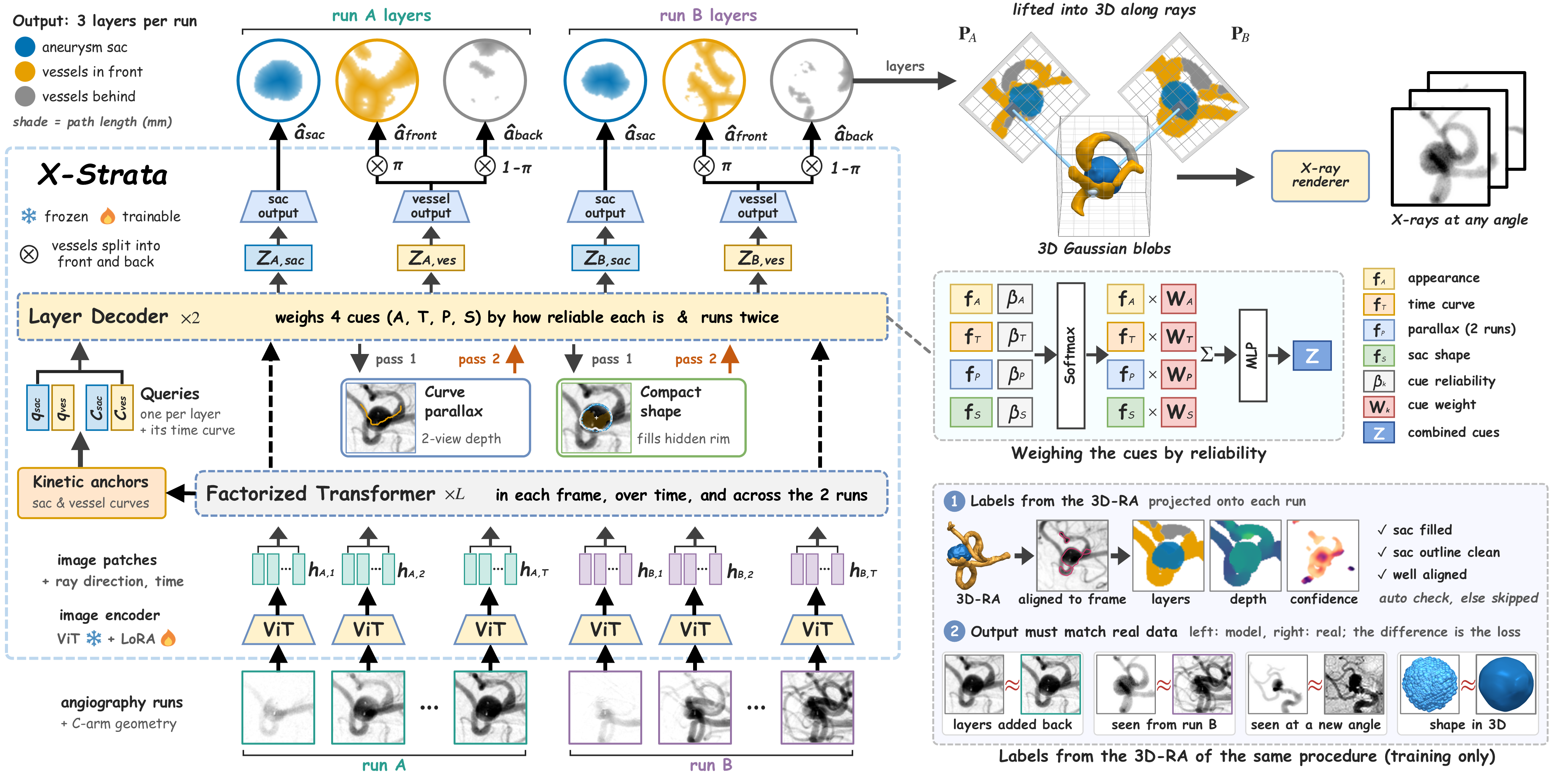}
\caption{\textbf{Two-pass architecture of \method.}
A ray-aware encoder processes cine frames and their C-arm geometry.
Kinetic anchors initialize the layer time curves.
The first-pass maps provide vessel centerlines and the visible sac rim for estimating parallax and shape cues.
The second pass weights appearance, time, parallax, and shape to estimate layer path lengths and front-vessel probabilities.
The layer maps are lifted into a shared 3D Gaussian representation for rendering.
During training, projected labels from same-session 3D rotational angiography (3D-RA) supervise eligible pre-deployment runs. A pixel-wise gate controls path-length supervision, and depth-order labels are weighted separately.}
\label{fig:method}
\end{figure*}

\paragraph{Image formation.}
Consider a ray that crosses the sac and a vessel. In each cine frame, the pixel records the sum of their contrast attenuation. For run $n$, the DICOM projection $P_n$ defines the ray through pixel $\vu$. Integrating the attenuation field of layer $\ell$ along this ray gives its effective path length $a_{\ell,n}$, where $\ell$ ranges over the sac and the front and behind vessels, $\mathcal K=\{\mathrm s,\mathrm f,\mathrm b\}$. A run contains one contrast injection, so each run has its own contrast curves $c_{\ell,n}$. A frame is the sum over layers of each path length times its curve, shifted by a local arrival delay $\tau_{\ell,n}$,
\begin{equation}
I_n(\vu,t)=\sum_{\ell\in\mathcal K}a_{\ell,n}
 c_{\ell,n}(t-\tau_{\ell,n})+\eta .
\label{eq:model}
\end{equation}
The right-hand side suppresses the dependence on $\vu$, and $\eta$ accounts for noise in the log-subtracted frame. A positive path length identifies a layer's support even at a shared pixel. We measure vessel depth relative to the sac along the ray, increasing away from the source, so the same vessel may change sides relative to the sac across views.

\paragraph{P1. A single parallel view leaves depth ambiguous.}
Exchange the positions of two structures while keeping each structure's path length, contrast curve, and delay fixed. Their order along the ray changes, and the sum in \cref{eq:model} stays the same. More generally, redistributing mass along the ray preserves the frames as long as the layer integrals and kinetics stay fixed, so one parallel projection can admit many depths. Perspective magnification can supply additional information outside this approximation. When the front and behind vessels share a contrast curve and have equal delays, they contribute only through their total path length, so a cine reconstruction can match even if their order is wrong. We therefore keep both signs in the depth distribution (\cref{sec:heads}) and train the order with a separate loss (\cref{sec:supervision}).

\paragraph{P2. Separating the contributions.}
At one pixel, suppose we know two nonnegative, unit-norm curves after applying their delays. Fitting their weighted sum to the measured samples estimates the two path lengths. If the curves meet at angle $\theta$, their inner product is $\cos\theta$, and the smaller Gram eigenvalue is $1-\cos\theta$. Noise along the least constrained direction changes the least-squares path lengths by a factor of up to
\begin{equation}
\kappa(\theta)=\frac{1}{\sqrt{2}\sin(\theta/2)},\qquad \theta>0 .
\label{eq:condition}
\end{equation}
Identical curves constrain only the sum of their coefficients, and the amplification falls as the curves separate. Reciprocally scaling a curve and its path length preserves the prediction, so projected path-length labels set the scale. We estimate the curve separation within each injection (\cref{sec:anchors}) and give the cine signal more weight as the curves separate (\cref{sec:fusion,sec:supervision}).

\paragraph{P3. Testing the depth sign in a second view.}
Given an anchor run and a partner run, take two candidate vessel locations with the same anchor-run coordinate and opposite offsets $\pm|\Delta z|$. Rotating the view by $\varphi$ moves their coordinates in the partner run apart by $2|\Delta z|\,|\sin\varphi|$. With equal priors and Gaussian localization errors of standard deviation $\sigma$ in that coordinate, we choose the hypothesis closest to the observed position, and a mistake occurs when the error crosses the midpoint. The probability of remaining on the correct side follows from the standard normal CDF $\Phi$,
\begin{equation}
q_{\mathrm{ord}}
=\Phi\!\left(\frac{|\Delta z|\,|\sin\varphi|}{\sigma}\right).
\label{eq:parallax}
\end{equation}
With zero projected separation, this test reduces to chance. Larger separation provides evidence for centerline matching, which we constrain using first-order projection consistency~\cite{natterer2001} (\cref{sec:ecp}). For the sac contour, the visible rim gives direct boundary evidence (\cref{sec:shape}).

\section{Method}
\label{sec:method}

The first pass of \method estimates projected sac and vessel layers from the cine frames and their rays (\cref{fig:method}). From these maps we extract vessel centerlines and locate the exposed sac rim. The second pass uses these observations to revise the layer maps and infer signed depth. The resulting path lengths and contrast curves render the cine, and the layers are lifted along their rays into a shared 3D Gaussian representation.

\subsection{Encoding a cine run and a view pair}
\label{sec:tokens}
\label{sec:anchors}

We encode each frame with DINOv2 ViT-B/14~\cite{dinov2_2024} and LoRA~\cite{lora2022}. A time embedding and a Pl\"ucker ray embedding~\cite{lfn2021,gslrm2024} locate each $14{\times}14$ patch in the cine and along the X-ray beam. Each block attends within a frame and then divides the tokens into two streams. The kinetic stream compares frames within an injection, while the geometry stream pools each patch location over time before exchanging information across runs. This separation keeps a change in bolus timing from serving as a cross-view correspondence. Cross-run attention uses sigmoid weights so multiple layers can draw evidence from one partner ray~\cite{epipolar2020,sigmoidattn2025}.

The layer curves start from pixels dominated by one layer, following the role of pure samples in separable NMF~\cite{gillis2014}. A purity head ranks pixels by their estimated single-layer content, and the $M$ highest-ranked pixels of each type initialize the corresponding curve query. We then alternate path-length and curve updates for $R$ unrolled nonnegative least-squares steps. Each curve is a weighted sum of gamma variates with a run-specific shift, and combining the shift with a local delay yields the arrival time in \cref{eq:model}. The continuous curves permit sub-frame delays. Front and behind vessels share a curve within an injection and may have different local delays.

\subsection{Centerline search with a moment anchor}
\label{sec:ecp}

To test the parallax evidence of P3, we thin the first-pass vessel map into branch-free segments. For node $i$ at coordinate $x_i$ on epipolar plane $h_i$ of the anchor run, we search $D$ nonzero signed offsets $z$ from the sac. Comparing these candidates in the partner run requires the depth of the sac, so we first compute a plane-level reference. We take separate path-length-weighted centroids for the sac and total vessels in each view. On plane $h$, let $\bar x_h^{\,n,\ell}$ and $\bar x_h^{\,m,\ell}$ denote these centroids in the anchor and partner runs. Under parallel projection of the same layer mass, subtracting the anchor centroid times $\cos\varphi$ from the partner centroid isolates the depth term. Dividing by $\sin\varphi$ recovers the mean layer depth in the frame of the anchor run,
\begin{equation}
\bar z_h^{\,\ell}
=\frac{\bar x_h^{\,m,\ell}-\bar x_h^{\,n,\ell}\cos\varphi}{\sin\varphi}.
\label{eq:moment}
\end{equation}
We apply this relation when both centroids are defined and $\sin\varphi\ne0$. Subtracting the sac moment from the total-vessel moment gives the moment anchor $\Delta\bar z_h$, which acts as a soft constraint on individual nodes.

Adding the sac moment to a candidate $z$ gives its depth in the anchor run. Its partner coordinate $x'_i(z)$ is the sum of the lateral term $x_i\cos\varphi$ and the depth term $(\bar z^{\,\mathrm s}_{h_i}+z)\sin\varphi$, and we read the soft centerline likelihood $\Lambda_m$ at that position. The score favors candidates that land on a partner centerline and penalizes those far from the moment anchor, with penalty scale $s_a$,
\begin{equation}
U_i(z)=\log\Lambda_m\big(h_i,x'_i(z)\big)
       -\frac{(z-\Delta\bar z_{h_i})^2}{2s_a^2}.
\label{eq:unary}
\end{equation}
Adjacent nodes are linked by a cost of $\nu$ times their absolute depth difference. Forward--backward inference combines this continuity cost with the candidate scores to return a marginal $q_i(z)$ at each node. At a vessel pixel, we sum the negative-depth probabilities of its nearest node and add a learned correction in log-odds space to obtain $\pi$. With a single run, $\pi$ comes from a learned front-class logit.

Several depth hypotheses can use the same partner pixel, which relaxes the one-to-one constraint of opaque stereo~\cite{ohta1985}. The soft matching scores and centroids also give depth supervision a path to the layer maps in both views for a fixed search grid.

\subsection{Visible-rim completion and cue fusion}
\label{sec:shape}
\label{sec:fusion}

The covered part of the sac has no direct boundary evidence, so we complete it from the exposed rim. We fit a star-convex contour~\cite{stardist2018,nosrati2015} with three Fourier harmonics to the rim, and weighted least squares gives zero weight to rim points covered by vessels. Rasterizing the fit supplies a shape cue for the covered region. With two runs, a degree-two spherical-harmonic surface fitted to the two rims initializes the sac Gaussians.

The second pass combines appearance, kinetic, parallax, and shape features. Each learned logit $\ell_k$ scores how well cue $k$ matches the layer query, and a reliability $r_k$ scales its unnormalized attention. Appearance has unit reliability. For kinetics, $1-\cos\hat\theta$ increases as the first-pass curves separate (\cref{eq:condition}). For shape, we use the fraction of exposed rim. For parallax, we square $|\Delta\hat z\sin\varphi|/\sigma_{\mathrm P}$ and cap it at one, using the first-pass depth offset $\Delta\hat z$ and a positive scale $\sigma_{\mathrm P}$. Normalizing the scaled scores across cues gives
\begin{equation}
w_k=\frac{r_k\exp(\ell_k)}{\sum_j r_j\exp(\ell_j)} .
\label{eq:bias}
\end{equation}
The decoder combines the cue features with weights $w_k$. With no eligible partner, zero parallax reliability removes that term, while appearance keeps the denominator positive. In practice we apply softmax to $\ell_k+\beta_k$, with $\beta_k=\log r_k$ for cues with positive reliability.

\subsection{Signed depth and rendering}
\label{sec:heads}

We use $\pi$ to choose the probability mass on each side and predict the spread of depths within each side. To construct $f_+$, we restrict a Gaussian with location $d\geq0$ and scale $s_z>0$ to the positive half-axis and divide by $\Phi(d/s_z)$. Reflecting it about zero gives $f_-(z)=f_+(-z)$. The front and behind probabilities weight the two components,
\begin{equation}
p(z)=\pi f_-(z)+(1-\pi)f_+(z).
\label{eq:depth}
\end{equation}
Integrating over negative depths recovers $\Pr(Z<0)=\pi$, even at $d=0$, because all mass of $f_-$ stays on that side. The same probability splits the vessel layer. For total vessel path length $a_{\mathrm v}$, we assign $\pi a_{\mathrm v}$ to the front layer and $(1-\pi)a_{\mathrm v}$ behind, which preserves the total for every $\pi$. The compositor inserts these path lengths and the local delays into \cref{eq:model}.

We lift the layers along their rays at the predicted depths, with Gaussian extents set by the layer path lengths. The selected runs contribute to a shared spatial density, while contrast curves remain specific to their injections. The R$^2$-Gaussian line integral~\cite{r2gaussian2024} renders this density in another view.

\subsection{Supervision from registered 3D-RA}
\label{sec:supervision}

The label protocol segments the sac and vessels in same-session 3D-RA and refines a six-degree-of-freedom pose for each run by maximizing the cross-correlation of the projected vessel tree with the peak cine frame. The registered volume supplies amodal support, path length, and signed depth in eligible pre-deployment runs.

Two geometric sac segmentations provide a check on the reference boundary. The protocol excludes rays that intersect voxels on which the two masks disagree, and it gives a procedure zero label weight if the 3D Dice of those masks falls below $0.85$. Additional checks screen inadequate opacification and poor registration. Rotational frames, illustrated alongside vessel alignment in Supplementary Fig.~S1, can supply rendering targets. Curve mixtures used in pretraining draw on training patients.

Registration quality weights the reference at pixel $\vu$ in run $n$ by $\omega_n=r_0^2/(r_0^2+r_n^2)$, using the local registration residual $r_n$ and a positive scale $r_0$. Rejected or unavailable labels receive zero weight. Temporal conditioning sets the weight on the cine. We estimate the relative perturbation $\epsilon_{0,n}$ associated with a one-frame shift, multiply it by the amplification $\kappa_n$ from \cref{eq:condition}, and convert this sensitivity to a temporal score $u_n=\exp(-\kappa_n\epsilon_{0,n})$, taking $u_n=0$ at a zero curve angle. The gate $g_n=u_n/(u_n+\omega_n)$ controls the weight on cine reconstruction when a valid label is available. With no label, we use $g_n=1$.

At each pixel, $\delta_n^{\mathrm{mix}}$ compares the predicted sac and total-vessel path lengths with their projected references, and $e_n$ compares the rendered cine with the measured frames. The term $\delta_n^{\mathrm{lab}}$ collects the semantic loss and binary cross-entropy on the order probability $\pi$, omitting terms without a reference. The training objective is
\begin{equation}
\mathcal L=\sum_{n,\vu}
 \big[\omega_n\delta_n^{\mathrm{lab}}+(1-g_n)\delta_n^{\mathrm{mix}}
      +g_n e_n\big]+\lambda\mathcal L_{\mathrm{3D}},
\label{eq:loss}
\end{equation}
where $\mathcal L_{\mathrm{3D}}$ combines surface and occupancy losses for eligible 3D references. As the curves become easier to separate, the gate shifts the path-length supervision toward the observed cine. The order loss retains its registration weight, including in a single view.

\paragraph{Inference.}
The detector sees only the runs allowed as input. With one run, it locates the crop from that run alone. With multiple runs, we can triangulate a common crop center from the allowed views and use independent 2D crops when the geometry is degenerate. Inference uses the selected cine runs and their geometry, and held-out 3D-RA and expert-corrected masks are reserved for scoring.

\section{Evaluation}
\label{sec:experiments}
\label{sec:setup}
\begin{table*}[t]
\centering
\caption{\textbf{Amodal layer decomposition across cohorts.} Squares denote appearance (\cueon{cA}), time (\cueon{cT}), parallax (\cueon{cP}), transmission physics (\cueon{cX}) and explicit shape (\cueon{cS}). O-Dice and DOA are defined in \cref{sec:experiments}. MAE is in mm. $^\dagger$Training-free solver using nnU-Net masks. Patients receive equal weight after averaging runs and seeds. \textbf{Bold} and \underline{underlining} mark the best and next distinct displayed values within each fixed input count. Dashes indicate ineligible endpoints.}
\label{tab:main}
\footnotesize\setlength{\tabcolsep}{6.8pt}\renewcommand{\arraystretch}{1}
\begin{tabular}{@{}lcrrrrrrr@{}}
\toprule
& & \multicolumn{5}{c}{\shortstack{\textbf{Internal test}\\Centers A and B ($n=\sNtestInt{}$)}} & \multicolumn{2}{c}{\shortstack{\textbf{External test}\\Center C ($n=\sNtestExt{}$)}} \\
\cmidrule(lr){3-7}\cmidrule(l){8-9}
Method & Cues & Sac Dice$\uparrow$ & O-Dice$\uparrow$ & Vessel Dice$\uparrow$ & DOA$\uparrow$ & MAE$\downarrow$ & O-Dice$\uparrow$ & DOA$\uparrow$ \\
\midrule
\multicolumn{9}{@{}l}{\textbf{One input run}} \\
nnU-Net 2D+t & \cueon{cA}\cueon{cT}\cueoff\cueoff\cueoff & \second{.699} & \second{.607} & .692 & -- & -- & .533 & -- \\
SAM 2 & \cueon{cA}\cueon{cT}\cueoff\cueoff\cueoff & .692 & .598 & .687 & -- & -- & \best{.569} & -- \\
SAM 3 & \cueon{cA}\cueon{cT}\cueoff\cueoff\cueoff & .688 & .595 & .682 & -- & -- & .546 & -- \\
Diffusion-VAS & \cueon{cA}\cueon{cT}\cueoff\cueoff\cueoff & .691 & .598 & \second{.696} & -- & -- & .536 & -- \\
Depth Anything V2 + 2 layers & \cueon{cA}\cueoff\cueoff\cueoff\cueoff & .654 & .561 & .685 & \second{.501} & \second{5.57} & .529 & \best{.517} \\
Ellipse (visible sac)$^\dagger$ & \cueon{cA}\cueoff\cueoff\cueoff\cueon{cS} & .676 & .579 & -- & -- & -- & \second{.559} & -- \\
\rowcolor{oursbg} \method{} & \cueon{cA}\cueon{cT}\cueoff\cueon{cX}\cueon{cS} & \best{.783} & \best{.641} & \best{.716} & \best{.528} & \best{5.20} & .552 & \second{.496} \\
\addlinespace[3pt]
\multicolumn{9}{@{}l}{\textbf{Two input runs}} \\
VGGT & \cueon{cA}\cueoff\cueon{cP}\cueoff\cueoff & .744 & .659 & \second{.694} & .644 & 4.63 & .589 & .609 \\
Epipolar transformer & \cueon{cA}\cueon{cT}\cueon{cP}\cueoff\cueoff & .748 & .655 & .679 & .648 & 4.50 & .595 & .614 \\
DIF-Net $\to$ project & \cueon{cA}\cueoff\cueon{cP}\cueon{cX}\cueoff & .765 & .674 & .690 & .651 & 4.33 & .606 & .623 \\
C$^2$RV $\to$ project & \cueon{cA}\cueoff\cueon{cP}\cueon{cX}\cueoff & \second{.767} & \second{.679} & .693 & .662 & \second{4.14} & \second{.630} & .626 \\
Ellipsoid (2 runs)$^\dagger$ & \cueon{cA}\cueoff\cueon{cP}\cueoff\cueon{cS} & .665 & .577 & -- & -- & -- & .519 & -- \\
Centerline stereo$^\dagger$ & \cueon{cA}\cueoff\cueon{cP}\cueoff\cueoff & -- & -- & -- & .680 & 4.43 & -- & .637 \\
Moment-anchored centerline stereo$^\dagger$ & \cueon{cA}\cueoff\cueon{cP}\cueon{cX}\cueoff & -- & -- & -- & \second{.691} & 4.30 & -- & \second{.652} \\
\rowcolor{oursbg} \method{} & \cueon{cA}\cueon{cT}\cueon{cP}\cueon{cX}\cueon{cS} & \best{.807} & \best{.693} & \best{.700} & \best{.703} & \best{3.21} & \best{.641} & \best{.659} \\
\addlinespace[3pt]
\multicolumn{9}{@{}l}{\textbf{All available input runs}} \\
\method{} & \cueon{cA}\cueon{cT}\cueon{cP}\cueon{cX}\cueon{cS} & .810 & .700 & .700 & .707 & 3.17 & .655 & .668 \\
\bottomrule
\end{tabular}
\end{table*}

We evaluate sac recovery under overlap, vessel order, and the contribution of each component.

\subsection{Cohort and input budgets}
The reported DSA cohort comprises \sNdsa patients at three centers. The treated aneurysm subset contains \sNcases patients, including \sNfd treated with flow diverters and \sNweb with intrasaccular devices. Each patient contributes one index aneurysm and at most one labeled treatment procedure. After excluding patients without same-session 3D-RA or an eligible pre-deployment run, with labels that failed the checks in \cref{sec:supervision}, or with severely degraded image quality, the evaluation pool contains \sNproc patients and \sNrunsEval cine runs. Centers A and B provide \sNtrain training, \sNval validation, and \sNtestInt internal-test patients (\sNrunsInt test runs), and Center C contributes \sNtestExt external-test patients in \sNrunsExt runs (Supplementary Fig.~S2). Patients are partitioned before preprocessing and model selection, and the full configuration was selected on the validation patients. Expert corrections to the 3D-RA references are reserved for scoring after prediction. Flagged pixels are excluded from scoring, and tied reference depths are excluded from order scoring.

An input budget fixes the runs a method may use: one run, two runs, or all available runs. The anchor is a pre-deployment treatment cine run, and one-run methods use only the anchor for both cropping and inference. For a two-run input, the partner comes from the same procedure and maximizes $|\sin\varphi|$ over eligible, nondegenerate pairs, with ties resolved by acquisition order, following the angular dependence in \cref{eq:parallax}. Methods at the same input budget share anchor runs, eligible partners, and crop rules.

\subsection{Comparators and endpoints}
Two kinds of comparison answer different questions. End-to-end systems compare the complete mapping from cine to layers, so their depth scores reflect both segmentation and depth estimation. We score the reconstruction baselines after projecting their 3D sac estimates into the anchor view, so their sac scores refer to the same detector region used for systems that predict layer masks directly. Matched-mask stereo holds the 2D+$t$ nnU-Net masks, crops, and run pairs fixed and compares two centerline solvers, with the moment penalty as the intervention. Projected C$^2$RV is the primary two-run sac contrast, and moment-anchored centerline stereo is the primary order contrast.

\Cref{tab:main} shows six one-run and seven two-run baseline configurations from rosters of 31 and 14, including nnU-Net~\cite{nnunet2021}, SAM 2 and SAM 3~\cite{sam2_2024,sam3_2025}, Diffusion-VAS~\cite{dvas2025}, Depth Anything V2~\cite{dav2_2024}, VGGT~\cite{vggt2025}, DIF-Net~\cite{difnet2023}, and C$^2$RV~\cite{c2rv2024}. The numerical appendix lists the full set. We count each configuration once per budget and treat seeds as repetitions.

\paragraph{Sac support and depth order.}
Overlap-region sac Dice (O-Dice) isolates the region where sac recovery depends on separating the layers. It restricts the predicted and reference sac masks to a two-pixel dilation of the reference vessel mask and uses the usual Dice coefficient on that region when either mask is nonempty. Balanced depth-order accuracy (DOA) averages the recalls of the front and behind classes in the overlap region and requires both classes in the reference. MAE measures absolute vessel-depth error in millimeters. Supplementary Fig.~S3 shows projected layer examples. For Dice and DOA, we average runs within each patient and seed before averaging seeds within the patient, so patients with more runs do not receive greater weight in the cohort mean.

\begin{figure*}[t]
\centering
\includegraphics[width=\linewidth]{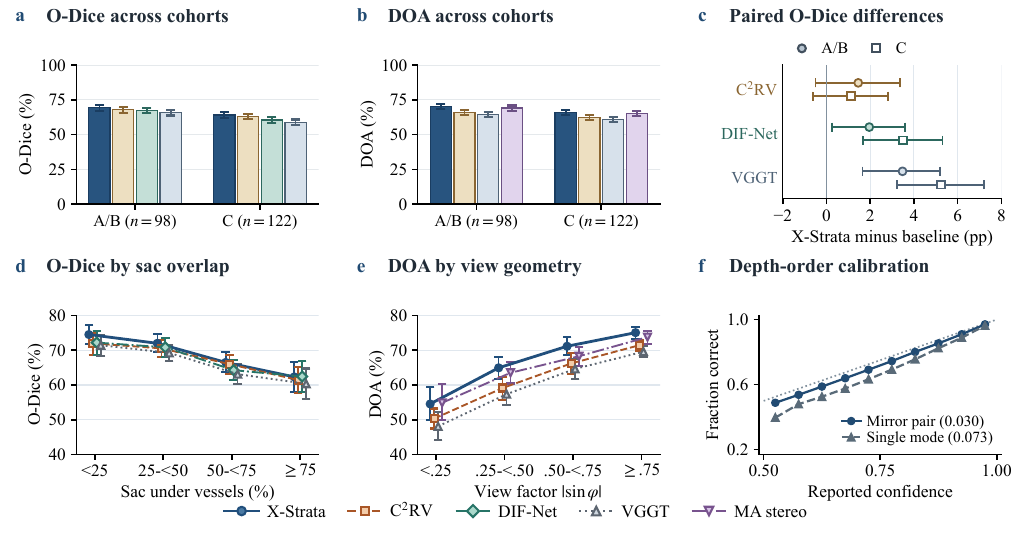}
\caption{\textbf{Two-run comparisons across cohorts.}
(a, b) Cohort means for O-Dice and balanced depth-order accuracy (DOA).
(c) Patient-paired O-Dice differences for \method against projected C$^2$RV, projected DIF-Net, and VGGT.
(d, e) Internal O-Dice by sac overlap and DOA by the view factor $|\sin\varphi|$.
Whiskers in (a--e) give pointwise 95\% intervals from 3,000 patient bootstrap draws. A patient can contribute to several bins in (d, e).
(f) Reliability curves pool eligible overlap pixels across seeds. Parentheses give the mean per-seed expected calibration error.}
\label{fig:paired_results}
\end{figure*}

\paragraph{Uncertainty and calibration.}
For each paired contrast, we take the mean of the patient-level differences among patients scored by both configurations. The 95\% pointwise intervals use 3,000 paired patient bootstrap draws. The signed-rank comparisons apply Holm correction within each cohort, endpoint, and input budget, using the eligible configurations in that family.

A front-class probability of at least $0.5$ predicts the front class, and the confidence assigned to the predicted class is $\max(\pi,1-\pi)$. Expected calibration error (ECE) groups eligible pixels into ten equal-width bins from $0.5$ to $1$ and averages the absolute gap between confidence and accuracy, weighted by the number of pixels in each bin. We pool pixels within seed before computing ECE, then average across seeds. The reliability plot pools counts across seeds.

\paragraph{Kinetic endpoint.}
The controlled-mixture protocol pairs sac and vessel curves on the same acquisition time grid and retains the unmixed sac curve as the reference. Time-to-peak (TTP) is the earliest time of maximum sac contrast on the sampled grid, and we report its mean absolute error at the patient level. Supplementary Fig.~S4 illustrates the contrast phases.

\subsection{Comparisons by input budget}
Each row of \cref{tab:main} summarizes its eligible inputs, so contrasts between budgets are descriptive, while two-run contrasts pair patients within the budget (Supplementary Fig.~S5). With one run, \method has higher O-Dice than the displayed comparators in A/B, and SAM 2 has the highest displayed O-Dice in C. With two runs, \method has the highest O-Dice and DOA among the eligible baselines in both cohorts.

\paragraph{Sac recovery.}
Against projected C$^2$RV, the paired O-Dice difference in \cref{fig:paired_results}c is \newIntODDelta{} percentage points in A/B (95\% interval \newIntODLo{} to \newIntODHi{}) and \newExtODDelta{} points in C (\newExtODLo{} to \newExtODHi{}), and both intervals include zero. Across the 12 eligible baselines, the Holm-adjusted $p$-value is below $0.05$ for 10 in A/B and 11 in C. The O-Dice point estimates for \method and projected C$^2$RV decrease across higher overlap bins in A/B (\cref{fig:paired_results}d). In 3D, median absolute sac-volume error for \method is \newIntVol\% in A/B and \newExtVol\% in C, and median sac Chamfer distance is $0.34$\,mm and $0.48$\,mm. On controlled mixtures, the internal two-run summary covers \sNrunsInt runs from \sNtestInt patients, with a mean absolute TTP error of \newTTP{}\,s on a $1/6$\,s sampling grid.

\paragraph{Vessel order.}
DOA is $0.49$ to $0.53$ across the four eligible one-run methods, consistent with P1, and $0.58$ to $0.70$ across the two-run methods. Against moment-anchored centerline stereo, the paired DOA difference is \newIntDOADelta{} points in A/B (95\% interval \newIntDOALo{} to \newIntDOAHi{}) and \newExtDOADelta{} points in C (\newExtDOALo{} to \newExtDOAHi{}). Only the A/B interval lies above zero, and its Holm-adjusted signed-rank $p$-value is $0.059$. Against the 11 end-to-end baselines, \method has higher DOA by at least $4.14$ points in A/B and $3.32$ points in C (Holm-adjusted $p<0.001$). DOA rises across the $|\sin\varphi|$ bins for both \method and the stereo comparator (\cref{fig:paired_results}e), consistent with P3, and reaches $0.75$ in A/B and $0.71$ in C for \method at $|\sin\varphi|\geq0.75$. Supplementary Fig.~S6 shows temporal, overlap, and view strata in both cohorts.

\subsection{Ablations of depth inference and layer recovery}
\begin{table}[t]
\centering
\caption{\textbf{Two-run ablations in the internal cohort.} Depth-only rows report DOA and expected calibration error (ECE). CD is median sac Chamfer distance in mm. TTP is patient-averaged absolute time-to-peak error in seconds on the mixture endpoint. Dashes indicate ineligible endpoints.}
\label{tab:ablation}
\footnotesize\setlength{\tabcolsep}{2.65pt}\renewcommand{\arraystretch}{1}
\begin{tabular}{@{}lrrrrr@{}}
\toprule
Variant & O-Dice$\uparrow$ & DOA$\uparrow$ & ECE$\downarrow$ & CD$\downarrow$ & TTP$\downarrow$ \\
\midrule
\rowcolor{oursbg} Full model (2 runs) & .693 & .703 & .030 & 0.34 & 0.144 \\
\addlinespace[2pt]
\multicolumn{6}{@{}l}{\textit{Depth ordering}} \\
Attention-only parallax & -- & .682 & .052 & -- & -- \\
Centerline parallax, frozen masks & -- & .694 & .050 & -- & -- \\
Without moment anchor & -- & .704 & .047 & -- & -- \\
With ordering constraint & -- & .674 & .054 & -- & -- \\
Single depth mode & -- & .710 & .073 & -- & -- \\
\addlinespace[2pt]
\multicolumn{6}{@{}l}{\textit{Shape, cues and temporal supervision}} \\
Without shape head & .693 & .710 & .047 & 0.35 & 0.127 \\
Without identifiability bias & .693 & .704 & .047 & 0.39 & 0.146 \\
Without kinetic anchors & .691 & .698 & .049 & 0.36 & 0.148 \\
Equal cue weights & .679 & .696 & .049 & 0.42 & 0.142 \\
Without temporal gate & .689 & .695 & .049 & 0.35 & 0.143 \\
\addlinespace[2pt]
\multicolumn{6}{@{}l}{\textit{3D output and registration}} \\
Without 3D lifting & .674 & .696 & .049 & -- & 0.151 \\
Labels from raw DICOM poses & .624 & .668 & .055 & 0.47 & 0.212 \\
\bottomrule
\end{tabular}
\end{table}

The ablations separate changes to the depth solver from changes to the full model. Attention-only parallax, removal of the moment anchor, the ordering constraint, and the single-depth variant use cached encoder features, masks, and crops, so these rows report DOA and ECE. The frozen-mask configuration also holds masks fixed but specifies a separately trained depth checkpoint. Changes to cue weighting, the temporal gate, the shape head, or the 3D loss use separate end-to-end checkpoints on the fixed training split.

\paragraph{Depth order and calibration.}
DOA falls when cross-run attention replaces centerline search and when an ordering constraint limits the allowed matches. These controls retain the second view and change how the solver uses it to score depths. \Cref{tab:ablation} also shows why order accuracy and calibration are separate endpoints. The single-depth configuration has higher DOA than the full depth head, while the full head has lower ECE (\newFullECE{} versus \newSingleECE{}). Removing the moment anchor leaves DOA close to the full model at table precision and raises ECE. In both cases a change in the depth distribution can affect confidence with little change to the thresholded decision. \Cref{fig:paired_results}f shows pooled reliability, and Supplementary Fig.~S7a,b shows individual seed traces and bin shares.

\paragraph{Layer recovery and labels.}
Equal cue weights give lower O-Dice and higher median surface error than reliability weighting. Removing the shape head leaves the displayed O-Dice unchanged and lowers TTP error. Raw-pose labels give lower O-Dice (Supplementary Fig.~S7c) and higher surface error than registered labels. These contrasts separate projected sac recovery, curve recovery, and sensitivity to label alignment.

\section{Discussion}
\label{sec:discussion}

\paragraph{What the layers add.}
An angiogram shows where contrast is present, but not which structure it belongs to or which vessel lies in front. \method estimates both from routine 2D cine runs at inference, which may be useful for patients who undergo only 2D angiography. Its order probabilities provide confidence scores. Mean per-seed ECE on eligible overlap pixels was \newFullECE{} in the internal cohort. Larger overlap regions contribute more pixels to this summary. Pixels from narrow and wide view pairs can also fall in the same confidence bin, even though depth order is harder to recover from narrow pairs. Comparing confidence and accuracy within each view bin could test whether the pooled agreement holds across view separations.

\paragraph{Geometry and failure modes.}
Order accuracy rises by about 20 points from the narrowest to the widest view bin in both cohorts, so \cref{eq:parallax} offers a practical rule for choosing runs. A vessel near the sac can still produce nearly identical partner coordinates for the two signs. At small $|\sin\varphi|$, independent centroid errors with variances $v_n$ and $v_m$ give the numerator of \cref{eq:moment} a variance of $v_m+v_n\cos^2\varphi$, amplified by $1/\sin^2\varphi$. Mask truncation and misregistration can shift the centroids at any angle, and magnification changes or movement between acquisitions can violate the assumed parallel projection and equal layer mass.

\paragraph{Comparison scope.}
Holding masks fixed isolates the solver's contribution, while the end-to-end comparison also reflects whether the model recovers each vessel, since a branch missing from either mask removes or weakens its match. The angle and overlap summaries are descriptive because a patient can contribute runs to several bins. Center C shares the scanner vendor, and acquisition changes can affect the bolus sampling and centerline localization underlying \cref{eq:condition,eq:parallax}.

\paragraph{Three-dimensional shape.}
Two views support a 3D sac estimate with median Chamfer distances of $0.34$ and $0.48$\,mm. Different surfaces can still produce the same two rims, and agreement in sac volume can coexist with local errors at the neck or boundary (Supplementary Fig.~S8).

\paragraph{Kinetics and clinical use.}
Mean TTP error on mixtures can fall below one frame interval because the mean includes zero-frame errors. All evaluated patients received 3D-RA, which scoring requires, and follow-up DSA differs in contrast delivery and temporal sampling. Extending \method to patients with only 2D angiography and relating recovered kinetics to residual-filling grades require a separate clinical reference.

\section{Conclusion}
\label{sec:conclusion}
{\interlinepenalty=10000
\method separates routine 2D cine runs into the complete sac and the vessels in front of and behind it, using contrast kinetics to estimate how much each layer contributes and parallax to constrain their order. On 220 held-out patients from three centers, it ranked first among 14 two-view baselines on overlap-region sac Dice and depth order. Because inference uses only 2D cine runs and their acquisition geometry, this approach may extend layered aneurysm and vessel information to patients who undergo only 2D angiography.
\par}

\paragraph{Use of AI tools.}
We used Codex (OpenAI) and Claude Code (Anthropic) during manuscript preparation and revision.
These tools assisted with drafting, literature searches, mathematical checks, and code development for numerical comparisons and figures.
The C-arm and patient illustration in \cref{fig:teaser} was generated with GPT-image-2 (OpenAI) and edited by the authors to depict the acquisition setup.

\FloatBarrier
{
    \small
    \bibliographystyle{ieeenat_fullname_appearance}
    \bibliography{refs_en}
}

\clearpage
\AddToShipoutPicture*{\put(0,0){\includegraphics[page=1,width=\paperwidth,height=\paperheight]{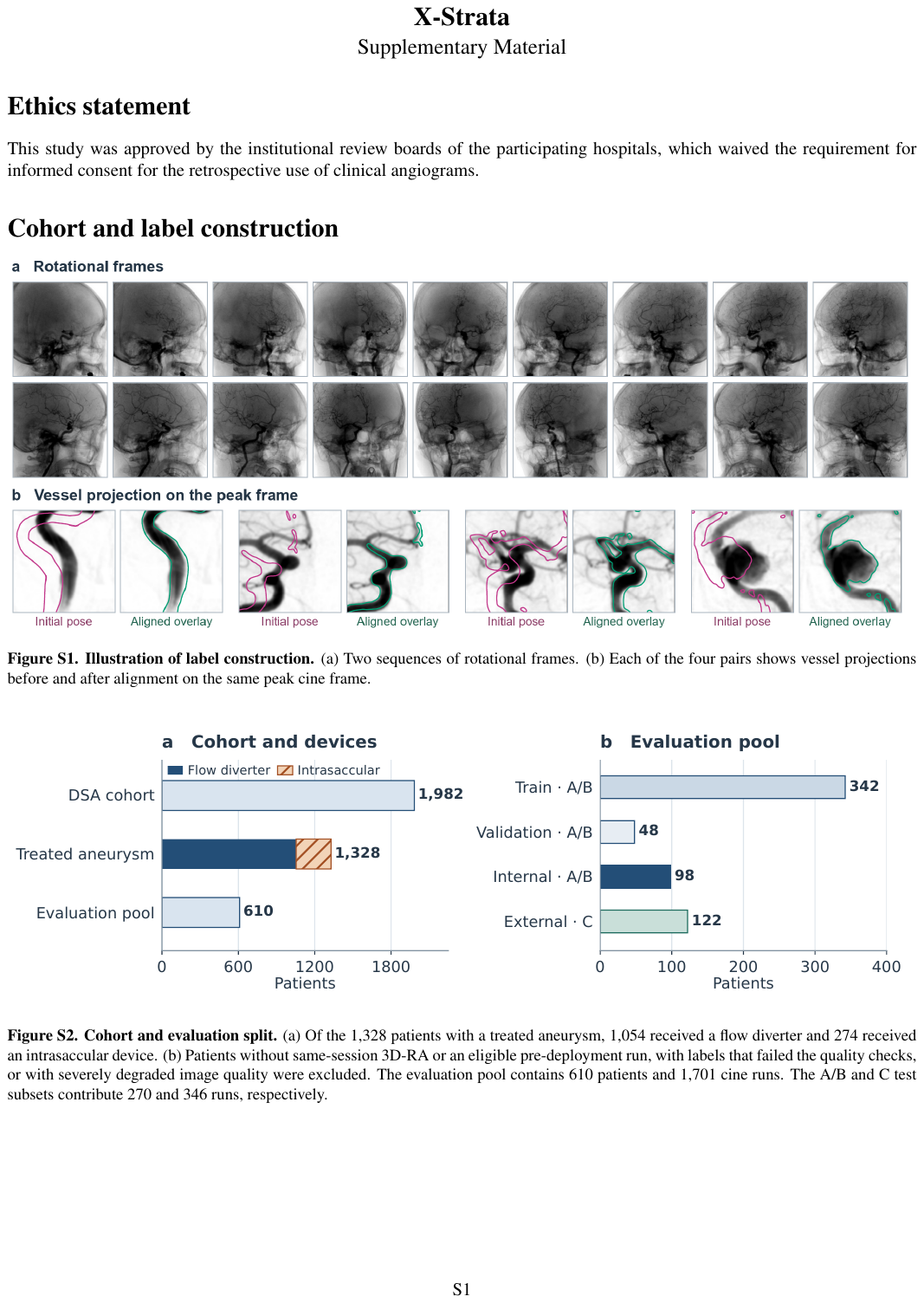}}}
\thispagestyle{empty}\mbox{}
\clearpage
\AddToShipoutPicture*{\put(0,0){\includegraphics[page=2,width=\paperwidth,height=\paperheight]{supplementary.pdf}}}
\thispagestyle{empty}\mbox{}
\clearpage
\AddToShipoutPicture*{\put(0,0){\includegraphics[page=3,width=\paperwidth,height=\paperheight]{supplementary.pdf}}}
\thispagestyle{empty}\mbox{}
\clearpage
\AddToShipoutPicture*{\put(0,0){\includegraphics[page=4,width=\paperwidth,height=\paperheight]{supplementary.pdf}}}
\thispagestyle{empty}\mbox{}
\clearpage
\AddToShipoutPicture*{\put(0,0){\includegraphics[page=5,width=\paperwidth,height=\paperheight]{supplementary.pdf}}}
\thispagestyle{empty}\mbox{}
\clearpage
\AddToShipoutPicture*{\put(0,0){\includegraphics[page=6,width=\paperwidth,height=\paperheight]{supplementary.pdf}}}
\thispagestyle{empty}\mbox{}
\clearpage
\AddToShipoutPicture*{\put(0,0){\includegraphics[page=7,width=\paperwidth,height=\paperheight]{supplementary.pdf}}}
\thispagestyle{empty}\mbox{}
\clearpage
\AddToShipoutPicture*{\put(0,0){\includegraphics[page=8,width=\paperwidth,height=\paperheight]{supplementary.pdf}}}
\thispagestyle{empty}\mbox{}
\clearpage
\AddToShipoutPicture*{\put(0,0){\includegraphics[page=9,width=\paperwidth,height=\paperheight]{supplementary.pdf}}}
\thispagestyle{empty}\mbox{}
\clearpage
\AddToShipoutPicture*{\put(0,0){\includegraphics[page=10,width=\paperwidth,height=\paperheight]{supplementary.pdf}}}
\thispagestyle{empty}\mbox{}
\end{document}